\documentclass[referee]{aa} 
\usepackage{graphicx}
\usepackage{txfonts}
\usepackage[round]{natbib}
\bibpunct{(}{)}{;}{a}{}{,} 
\usepackage{epsfig}
\usepackage{longtable}

\begin{document}

\title{Hunting for millimeter flares from magnetic reconnection in
  pre-main sequence spectroscopic binaries\thanks{Based on
    observations with the IRAM 30m telescope at Pico Veleta
    (Spain). IRAM is funded by the INSU/CNRS (France), the MPG
    (Germany) and the IGN (Spain).}}  \titlerunning{Millimeter
    flares from pre-main sequence spectroscopic binaries}

\author{
\'A. K\'osp\'al\inst{1}
\and
D. M. Salter\inst{1}
\and
M. R. Hogerheijde\inst{1}
\and
A. Mo\'or\inst{2}
\and
G. A. Blake\inst{3}}
\authorrunning{\'A. K\'osp\'al et al.}

\institute{
Leiden Observatory, Leiden University, PO Box 9513, 2300 RA Leiden,
The Netherlands\\ 
\email{kospal@strw.leidenuniv.nl}
\and
Konkoly Observatory of the Hungarian Academy of Sciences, PO Box 67,
1525 Budapest, Hungary 
\and
California Institute of Technology, Division of Geological and
Planetary Sciences, Mail Stop 150-21, Pasadena, CA 91125, USA 
}

\date{Received date; accepted date}


\abstract
{Recent observations of the low-mass pre-main sequence (PMS),
  eccentric spectroscopic binaries DQ\,Tau and V773\,Tau\,A reveal
  that their millimeter spectrum is occasionally dominated by flares
  from non-thermal emission processes. The transient activity is
  believed to be synchrotron in nature, resulting from powerful
  magnetic reconnection events when the separate magnetic structures
  of the binary components are briefly capable of interacting and
  forced to reorganize, typically near periastron.}
{We conducted the first systematic study of the millimeter variability
  toward a sample of 12 PMS spectroscopic binaries with the aim to
  characterize the proliferation of flares amongst sources likely to
  experience similar interbinary reconnection events. The source
  sample consists entirely of short-period, close-separation binaries
  that possess either a high orbital eccentricity ($e\,{>}\,$0.1) or a
  circular orbit ($e\,{\approx}\,$0).}
{Using the MAMBO2 array on the IRAM 30m telescope, we carried out
  continuous monitoring at 1.25\,mm (240\,GHz) over a
  4-night period during which all of the high-eccentricity binaries
  approached periastron. We also obtained simultaneous optical VRI
  measurements, since a strong link is often observed between
  stellar reconnection events (traced via X-rays) and optical
  brightenings.}
{UZ\,Tau\,E is the only source to be detected at millimeter
  wavelengths, and it exhibited significant variation ($F_{\rm
    1.25mm}$\,=\,87--179\,mJy); it is also the only source to undergo
  strong simultaneous optical variability
  ($\Delta{}R\,{\approx}\,$0.9\,mag). The binary possesses the largest
  orbital eccentricity in the current sample, a predicted factor in
  star-star magnetic interaction events. With orbital parameters and
  variable accretion activity similar to DQ\,Tau, the millimeter
  behavior of UZ\,Tau\,E draws many parallels to the DQ\,Tau model for
  colliding magnetospheres. However, on the basis of our observations
  alone, we cannot determine whether the variability is repetitive, or
  if it could also be due to variable free-free emission in an ionized
  wind.}
{UZ\,Tau\,E brings the number of known millimeter-varying PMS sources
  to 3 out of a total of 14 monitored binaries now in the
  literature. Important factors in the non-detection of the rest
  of our targets are the coarse time-sampling and limited millimeter
  sensitivity of our survey. We recommend that future studies
  concentrate on close-by targets, and obtain millimeter and optical
  data points with better temporal resolution.
}


\maketitle


\section{Introduction}

Recent observations of young stellar objects (YSOs) are challenging
the long-standing notion that the millimeter continuum emission
characterizing these objects is always dominated by the quiescent
thermal emission from passively heated circumstellar dust. Powerful,
transient millimeter flares attributed to synchrotron continua have
now been reported toward two embedded protostars in the Corona
Australis cloud \citep{choi2009}, one protostar in the Orion BN/KL
star-forming region \citep{forbrich2008}, the embedded YSO GMR-A in
Orion \citep{bower2003,furuya2003}, the classical T~Tauri star (CTTS)
DQ\,Tau in Taurus \citep{salter2008,salter2010}, and the weak-line
T~Tauri star (WTTS) V773\,Tau\,A, also in Taurus
\citep{massi2002,massi2006,massi2008}.

These millimeter flares are thought to be more powerful examples of
the prevalent, lower-energy radio activity observed toward YSOs, and
are not unlike millimeter flares occurring on the Sun
\citep{stine1988,white1992}. The emission is attributed to a
combination of gyrosynchrotron and synchrotron radiation powered by
magnetic reconnection events in the stellar coronae, which occur when
oppositely directed magnetic field lines interact \citep{bastian1998}.
The radio flare resulting from this magnetic activity is expected to
be accompanied by an X-ray flare of proportional luminosity according
to the Neupert effect \citep{neupert1968,gudel2002}. In this way,
large X-ray surveys are contributing to the characterization of the
magnetic activity during the T\,Tauri phase
\citep[e.g.][]{getman2005,gudel2007}, which represents the period
during the formation and main sequence life of a solar-type star when
magnetic activity levels are highest and when reconnection---and not
accretion---is believed to be the primary X-ray production mechanism
\citep{preibisch2005,stassun2006,stassun2007,forbrich2007,feigelson2007}.
The X-ray data confirm coronal activity analogous to that of the Sun,
but with luminosities 10$^3$--10$^5$ times higher \citep{testa2010}.
Uninterrupted, long-duration observing campaigns of star-forming
regions also document a once-a-week statistical occurrence of giant
X-ray flares, representing the most powerful events and an estimated
1\% of all flares \citep{favata2005,getman2008a,getman2008b}. If
magnetic activity is indeed the trigger, then these X-ray events are
most likely accompanied by radio events due to synchrotron emission
processes extending into the millimeter regime.

It is noteworthy that in the two best-studied millimeter flare cases,
both DQ\,Tau and V773\,Tau\,A are close-separation, eccentric,
pre-main sequence (PMS) binaries with similar orbital characteristics.
In addition, their flaring was recurring, leading the authors in the
latest study to conclude that DQ\,Tau could display excess millimeter
flux as much as 8\% of the time, with consistency near periastron. One
can thus speculate that in close binaries, in addition to the
single-star coronal activity described until now, two additional
magnetic activity scenarios might exist. An example of the first is
V773\,Tau\,A, where interbinary interactions due to chance alignments
of narrow extended coronal features, like helmet streamers, cause the
flares. The second binary-specific phenomenon is the current model for
DQ\,Tau where colliding dipolar magnetospheres represent a simple
geometric scenario for periodic events, with two primary determining
factors: a periastron approach smaller than twice the magnetospheric
radius ($R$\,$\sim$\,5\,$R_\star$) and a high eccentricity. In this
arrangement, the closed stellar magnetospheres must overlap near
periastron, but only temporarily.

Once again, X-ray studies can provide constraints on the coronal
extent of PMS stars, with the result that typical inferred loop
lengths are 4--20\,$R_\star$ for the most powerful outbursts. This is
much larger than any coronal structure observed toward more evolved
stars \citep{favata2005}, and consistent with the T\,Tauri stage being
the most magnetically active phase of star formation. In the case of
the WTTS binary system V773\,Tau\,A, two separate coronal structures
extending to $\geq$\,15\,$R_\star$ each are necessary to bridge the
interbinary gap \citep{massi2008}. Toward DQ\,Tau the derived loop
lengths from both X-ray and millimeter analyses are 5\,$R_\star$ in
height (Getman et al.~2010, submitted; Salter et al.~2010). If the
giant X-ray flare statistics correctly predict a common once-a-week
occurrence with consistently large loop lengths, then more interbinary
collisions might be expected in a number of close-separation binaries,
in addition to any single-star events that may occur. Binary systems
also occur frequently, representing 65\% or more of the local field
population in the middle of the main sequence
\citep{duquennoy1991}. The fraction increases to up to 75\% for the
population in the Taurus star-forming region
\citep{leinert1993,ghez1997,kohler1998,luhman2010}, suggesting that
more systems could start out as binaries. Thus, candidate
millimeter-variable systems are worthy of investigation.

To assess the proliferation of significant millimeter variability
among PMS binaries, we report here on a targeted millimeter
variability survey of 12 PMS spectroscopic binaries that are most
likely to experience millimeter flares based on predictions by the
current interbinary magnetic reconnection models, either following the
V773\,Tau\,A scenario and exhibiting strong flares at many orbital
phases, or exhibiting the DQ\,Tau phenomena showing flares with more
regularity around periastron. Since in both the DQ\,Tau and UZ\,Tau\,E
cases, optical brightenings are common near periastron due to periodic
accretion events \citep{jensen2007}, and because the optical light
curve of DQ\,Tau was found to mirror its millimeter flare activity in
both time and duration \citep{salter2010}, we complemented our
millimeter data with simultaneous optical monitoring of our targets.


\section{Observations}
\label{observations}

\subsection{Target selection}
\label{targetselection}

The study of PMS binaries is a relatively young field, due in large
part to the difficult and time-consuming nature of spectroscopic
observations, and sometimes further impeded by a complicated
circumstellar environment. To date, orbital parameters have been
published for only a few dozen young spectroscopic binaries. These
parameters include the \emph{necessary} selection criteria for the
present study: known eccentricity ($e$), orbital period ($P$),
projected semimajor axis ($a\,{\sin}\,i$), and the epoch of
periastron passage. From the available sources in the literature, we
selected objects that are observable from the northern hemisphere
(located primarily in Taurus and Orion), and that possess an orbital
period of less than 50\,days. The latter constraint means that the
periastron distances for these objects are in the range where
magnetospheric or coronal interactions are known to occur in similar
systems. In addition, shorter periods make it easier to observe large
numbers of binaries as they complete periastron passage around the
same time.

The resulting list of targets, along with their most important orbital
parameters, can be found in Table~\ref{tab:targetlist}. Our sample is
comprised of two separate groups. In the first group, we specifically
target short-period binaries of high orbital eccentricity
($e\,{>}\,$0.1) and small periastron separations similar to DQ\,Tau
and V773\,Tau\,A. In this half of the sample, the binary components
are suspected to undergo a change in their large-scale magnetospheric
topology as the binary separation distance varies greatly and rapidly
near periastron. This group is most likely to experience a merging of
the two magnetospheres near closest approach with a subsequent
detachment at larger separations -- much like the current picture for
DQ\,Tau \citep{salter2010}. Although we note that none of the targets
quite approach the exceptional combination of large eccentricity and
close approach as DQ\,Tau (with its $e$\,=\,0.556 and $d_{\rm
  min}$\,=\,13\,$R_{\odot}$). The second group serves as a ``control
sample'' and includes close-separation binaries with circular orbits
($e\,{\approx}\,$0). Since the binary separation in these systems
remains constant, there is no variable compression or relaxation of
the component magnetospheres throughout the orbit, and therefore the
global magnetospheric topology is relatively unchanged; reconnection
is not expected to occur except sporadically in chance collisions of
extended coronal features due to the stars' close proximity -- similar
to the current model for V773\,Tau\,A \citep{massi2008} -- or in rare
giant single-star flares.

\begin{table*}
\caption{Orbital parameters for the binaries. \label{tab:targetlist} }
\begin{center}
\begin{tabular}{lccccccc}
\hline
\hline
\noalign{\smallskip}
Name                          & $D$        &  $e$   &  $P$   &  $a$\,sin\,$i$ &  $i$         & $d_{\rm min}$ &  Ref. \\
                              & [pc]       &        & [days] & [AU]           & [$^{\circ}$] & [R$_{\odot}$] &       \\
\hline
\multicolumn{8}{c}{\textit{High-eccentricity binaries}}\\
\hline
\noalign{\smallskip}
\object{EK\,Cep}              & 164        & 0.109  & 4.43  & 0.077 & 89.3$^*$ & 14.7    & 1,2 \\
\object{UZ\,Tau\,E}           & 140        & 0.33   & 19.13 & 0.124 & 54       & 22.1    & 3 \\
\object{RX\,J0530.7$-$0434}   & 460        & 0.32   & 40.57 & 0.336 & 78.5     & $>$49.1 & 4,5 \\
\object{Parenago\,1540}       & 470        & 0.12   & 33.73 & 0.188 & $\cdots$ & $>$35.6 & 6 \\
\object{Parenago\,2494}       & 470        & 0.257  & 19.48 & 0.146 & $\cdots$ & $>$23.3 & 7 \\
\object{GG\,Ori}              & 438        & 0.222  & 6.63  & 0.116 & 89.2$^*$ & 19.4    & 8 \\
\hline
\multicolumn{8}{c}{\textit{Circular binaries}}\\
\hline
\noalign{\smallskip}
\object{RX\,J0350.5$-$1355}   & 450$^{**}$ & 0      & 9.28  & 0.115 & $\cdots$ & $>$24.7 & 4 \\
\object{V826\,Tau}            & 150        & 0      & 3.89  & 0.013 & 13       & 12.6    & 9 \\
\object{RX\,J0529.4$+$0041\,A}& 325        & 0      & 3.04  & 0.053 & 86.5$^*$ & 11.5    & 10 \\
\object{Parenago\,1802}       & 420        & 0.029  & 4.67  & 0.050 & 78.1$^*$ & 10.7    & 11 \\
\object{RX\,J0541.4$-$0324}   & 450$^{**}$ & 0      & 4.99  & 0.074 & $\cdots$ & $>$15.9 & 4 \\
\object{NGC\,2264\,Walk\,134} & 913        & 0      & 6.35  & 0.099 & $<$73    & $>$21.3 & 12,13 \\
\hline
\multicolumn{8}{c}{\textit{Binaries already known to exhibit flares}}\\
\hline
\noalign{\smallskip}
\object{DQ\,Tau}              & 140        & 0.556  & 15.80 & 0.053 & 23   & 12.9    & 14\\
\object{V773\,Tau\,A}         & 148.4      & 0.272  & 51.10 & 0.35  & 66.0 & 60.0    & 15\\
\hline
\end{tabular}
\end{center}
\textbf{Notes.} $e$ -- eccentricity; $P$ -- orbital period; $a$ --
semimajor axis; $i$ -- inclination; $d_{\rm min}$ -- separation during
periastron (or a lower limit for systems with an unknown
inclination); $^*$ -- indicates eclipsing binaries; $^{**}$ -- substituted average distance to Orion.\\
\textbf{References.}
1 -- \citet{gimenez1985}; 2 -- \citet{torres2010}; 3 --
\citet{jensen2007}; 4 -- \citet{covino2001};
5 -- \citet{marilli2007};
6 -- \citet{marschall1988};
7 -- \citet{reipurth2002};
8 -- \citet{torres2000};
9 -- \citet{reipurth1990};
10 -- \citep{covino2004};
11 -- \citet{stassun2008};
12 -- \citet{padgett1994};
13 -- \citet{baxter2009};
14 -- \citet{mathieu1997};
15 -- \citet{boden2007}. 
\end{table*}

\subsection{Observations and data reduction}
\label{datareduction}

\paragraph{Millimeter observations.} We obtained millimeter data with
the IRAM 30\,m telescope on Pico Veleta (Spain) between 17--21
November 2009. We used the 117-pixel Max-Planck Millimeter Bolometer
array (MAMBO2) in the standard ON/OFF mode, using a 35$''$ wobbler
throw, and making sure that the target fell on the most sensitive
pixel (pixel \#20) during each on-source exposure. One scan consisted
of four (on the first night) or eight (on the remaining three nights)
subscans of 60\,s exposure time. The MAMBO2 bandpass was centered at
$\lambda$\,=\,1.25\,mm ($\nu$\,=\,240\,GHz). At this wavelength, the
half power beamwidth is 11$''$ and the pixel spacing is 20$''$.

We monitored our targets for four consecutive nights, covering also
the periastron events for most of the high-eccentricity binaries. We
observed each target 1-3 times per night depending on weather
conditions. The atmospheric transmission at 1.25\,mm was usually good,
with the zenith atmospheric opacity being monitored with sky dips
every 1-2\,hours. Opacity was consistently found to be between
0.1--0.4, except for a few peaks on the night of November 19/20. Sky
noise levels were typically very low, except again for a few hours on
the night of November 19/20. Mars and Uranus were used for focusing
and gain calibration every 3-4 hours. Pointing was checked every
1-2\,hours using Mars, Uranus, or a nearby quasar. Occasionally, high
wind velocities during our observing run resulted in a lower pointing
precision for some of the scans. As a result, we have determined that
the absolute flux calibration for all data points is accurate to
within $\sim$20\%.

The data were reduced using the MOPSIC pipeline (developed by
R.~Zylka), which includes steps to remove the atmospheric emission
(using the two wobbler offset positions), to correct for extinction
due to atmospheric water vapor (using the skydips), to perform gain
calibration (using the calibration observations of bright sources with
known 1.25\,mm flux), and to calculate a noise-weighted average of the
subscans. The obtained flux densities usually have a root mean square
noise of 2-3\,mJy\,bm$^{-1}$, and are listed per source in
Table~\ref{tab:mm}.

\paragraph{Optical observations.} Simultaneous optical monitoring of
our targets was conducted from two telescopes: the 60/90/180\,cm
(aperture diameter/primary mirror diameter/focal length) Schmidt
telescope of the Konkoly Observatory (Hungary), and the 80\,cm
(primary mirror diameter) IAC-80 telescope of the Teide Observatory in
the Canary Islands (Spain). The Konkoly Schmidt telescope is equipped
with a 1536\,$\times$\,1024 pixel Photometrics AT\,200 CCD camera
(pixel scale: 1.03$''$) and a Bessel UBV(RI)$_{\rm C}$ filter
set. The Teide IAC-80 telescope is equipped with a
2048\,$\times$\,2048 pixel Spectral Instruments E2V 42-40
back-illuminated CCD camera `CAMELOT' (pixel scale: 0.304$''$) and a
Johnson-Bessel UBV(RI)$_{\rm J}$ filter set. Images with the Schmidt
telescope were obtained on 9 nights in the period between 13--26
November 2009, and with the IAC-80 telescope for 3 nights between
18--22 November 2009. With the IAC-80, only R band images were taken,
whereas with the Schmidt, R and I, and additionally for the brighter
stars, V, images were also obtained. One of our targets, UZ\,Tau\,E,
was observed with the IAC-80 telescope using VRI filters also between
25 October and 7 November 2009.

The images were reduced in IDL following the standard processing steps
of bias subtraction, dark subtraction (for the Photometrics camera on
the Schmidt telescope only) and flat-fielding. On each night, for each
target, images were obtained in blocks of 3 or 5 frames per filter.
Aperture photometry for the target and other field stars were
performed on each image using IDL's \textit{cntrd} and \textit{aper}
procedures. For the Schmidt images, we used a 5-pixel radius aperture
and a sky annulus between 10 and 15 pixels. For the IAC-80 images, the
aperture radius was 8 pixels and the sky annulus was between 30 and 40
pixels. In the case of RX\,J0529.4+0041, a hierarchical triple system,
the spectroscopic binary and the tertiary component were partially
resolved. As a result, we increased the Schmidt aperture to 10 pixels
and the IAC-80 aperture to 30 pixels, in order to make sure that the
aperture included the total flux from all three components. In the
case of UZ\,Tau, a hierarchical quadruple system, we obtained separate
photometry for UZ\,Tau\,E and UZ\,Tau\,W in the following way. The
small pixel scale and the good seeing at the IAC-80 telescope made it
possible to obtain separate photometry for both the E and the W
components, by using an aperture radius of 4 pixels. The results
indicated that the W component was constant within 0.02\,mag. For the
Schmidt images, we used an aperture radius of 10 pixels (encompassing
both UZ\,Tau\,E and W), and we subtracted the contribution of the W
component (calculated as the average magnitude measured on the IAC-80
images).

For the purpose of differential photometry, we selected a comparison
star for each of our targets. Our main selection criterium for the
comparison stars was constant brightness during our observing period
(compared to other stars in the field). The selected comparison stars
are listed in Table~\ref{tab:comp}. We calculated the magnitude
difference between our target and the comparison stars for each
frame. Then, these differential magnitudes were averaged for each
night, while their standard deviation was quadratically added to the
formal photometric uncertainty. Due to the different filter sets on
the two telescopes, the resulting magnitudes were slightly different,
especially in cases where the comparison star was much redder than the
target star. In these cases, we shifted the IAC-80 magnitudes by
0.03-0.08\,mag so that they overlap with the Schmidt magnitudes
obtained on the same nights. The resulting (shifted) values can be
found in Table~\ref{tab:opt}, and the light curves for the R filter
are plotted in the top panels of Figs.~\ref{fig:light_ecc} and
\ref{fig:light_circ}. The light curves for the other filters look very
similar, but often have fewer available data points.


\section{Results}
\label{results}


\begin{figure*}
\centering
\includegraphics[height=\textwidth,angle=90]{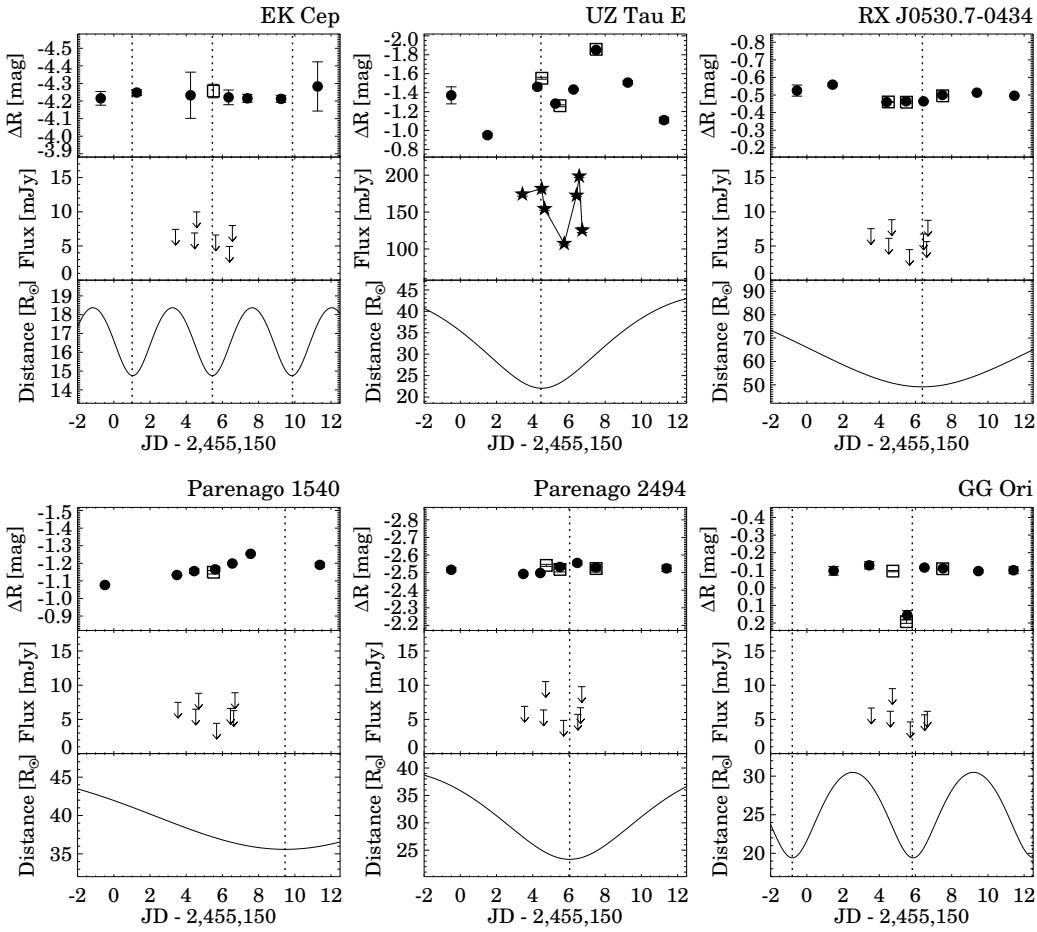}
\caption{Optical (top panels) and millimeter (middle panels) light
  curves for each of the high-eccentricity spectroscopic binaries, as
  well as separation distances for the binary components as a function
  of time (bottom panels). {\it Filled dots:} R$_{\rm C}$-band
  observations from the Konkoly Schmidt telescope in Hungary; {\it
    Open squares:} R$_{\rm J}$-band observations from the Teide IAC-80
  telescope in Spain; {\it Filled stars:} 1.25\,mm observations with
  the IRAM\,30\,m telescope in Spain; {\it Arrows:} 3$\sigma$ upper
  limits for the 1.25\,mm fluxes. Note that the millimeter fluxes of
  UZ\,Tau\,E contain an $\approx$20\,mJy contribution from
  UZ\,Tau\,W. A vertical line indicates the time of
  periastron. \label{fig:light_ecc}}
\end{figure*}

\begin{figure*}
\centering
\includegraphics[height=\textwidth,angle=90]{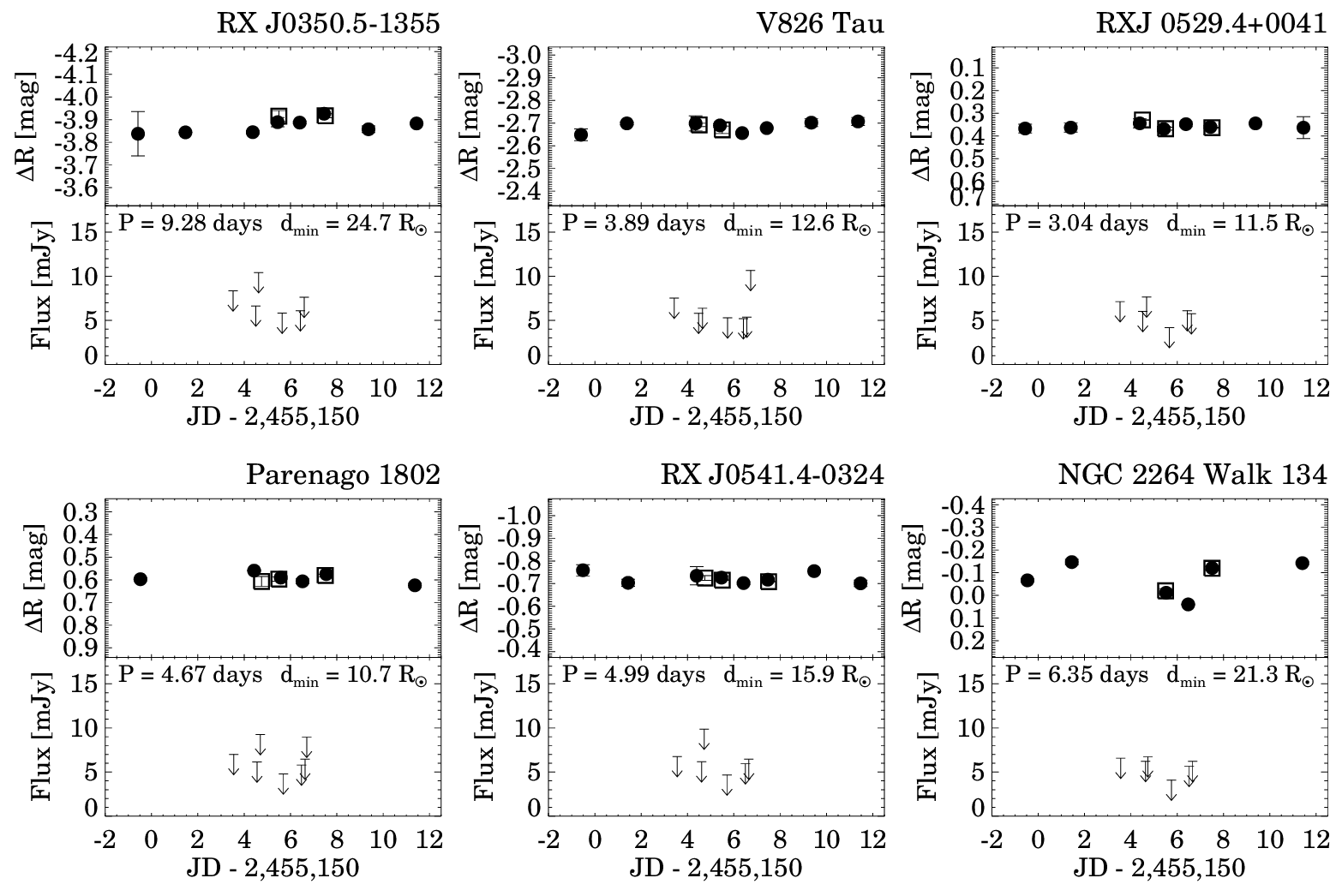}
\caption{Same as Fig.~\ref{fig:light_ecc}, but for the circular
  binaries. Separation distances, which are constant in these cases,
  are written in the lower panels.\label{fig:light_circ}}
\end{figure*}


In Figs.\,\ref{fig:light_ecc} and \ref{fig:light_circ} we show the
optical and millimeter light curves for all of our targets, as well as
the separation distances between the binary components, as a function
of time. The optical light curves indicate that most of our targets
showed variability at a certain level throughout our observing
campaign. By far the highest amplitude in our sample was displayed by
UZ\,Tau\,E, (with $\Delta$$R$\,$\approx$\,0.9\,mag), but
NGC\,2264\,Walk\,134 ($\Delta$$R$\,$\approx$\,0.2\,mag), and GG\,Ori
($\Delta$$R$\,$\approx$\,0.3\,mag) also varied significantly. The rest
of the sample were either constant (EK\,Cep, RX\,J0529+0041), or
showed slight variations of $<$0.1\,mag. While UZ\,Tau\,E and
NGC\,2266\,Walk\,134 show variability on a daily timescale, the light
curve for GG\,Ori exhibits only one dip, corresponding to an eclipse
shortly before periastron. The sparsely-sampled optical light curves
do not display any obvious periodicities in their own right, even with
the broader optical monitoring period of 12-14\,days and covering
multiple orbital periods for some sources. However, in many cases the
shapes and amplitudes may be consistent with rotational modulation of
the light curve due to stellar spots (RX\,J0530.7$-$0434,
Parenago\,1540, Parenago\,2494, RX\,J0350.5$-$1355, V826\,Tau,
Parenago\,1802, and RX\,J0541.4$-$0324). Thus, a brightening of an
eccentric source during periastron may be a chance coincidence. A more
detailed analysis of the optical data on a source-by-source basis is
given in Sect.~\ref{othersources}.

At millimeter wavelengths, only one of our targets, UZ\,Tau, was
detected. The other sources remained undetected with typical 3$\sigma$
upper limits between 4 and 11\,mJy. Apparently, UZ\,Tau is the only
source for which a potentially significant outer circumstellar disk
may still be present. The other sources in our sample are more similar
to V773\,Tau\,A, with little to no remaining circumstellar
material. We derive upper dust mass limits for the small-grain
population of $\le$\,3.0$\times$10$^{-5}$\,$M_\odot$ (0.003\,$M_\odot$
for the total gas plus dust mass) for the sources in Taurus and
$\le$\,2.5$\times$10$^{-4}$\,$M_\odot$ (0.025\,$M_\odot$ for the total
gas plus dust mass) for the sources in Orion, on average. To make
these dust mass estimates, we used the basic millimeter flux--mass
relation of \citet{dutrey1996} and take
$\kappa_v$\,=\,0.02\,g\,cm$^{-2}$ and $T_{\rm dust}$\,=\,45\,K. These
upper limits are consistent with the optical--near infrared SEDs for
these sources, which are usually fit well by a single-temperature
stellar photosphere in the available literature.

The fact that 10 out of 12 of our sources are WTTS is largely a
selection effect, due to the orbital parameters being easier to
determine when stellar photospheric lines are not obscured by
circumstellar material. In reality, the multiplicity ratio for CTTSs
and WTTSs in Taurus is essentially identical
\citep{kohler1998}. Interestingly, WTTSs tend to dominate the X-ray
sky in Taurus \citep{neuhauser1995,gudel2007}, suggesting that they
are more magnetically active. If WTTSs are also more likely to
experience the most powerful magnetic outbursts, our survey failed to
detect an example of such an event.

Instead, it is the CTTS UZ\,Tau that exhibits millimeter
variability. We detected UZ\,Tau at a level of 38--105$\sigma$ over
the course of our monitoring program, with its 1.25\,mm flux varying
between 107 and 199\,mJy in a period of less than 4 days, a clear
indication of a significant contribution of non-thermal emission
processes. The millimeter light curve in Fig.~\ref{fig:light_ecc}
shows two peaks: one at around periastron, and another one about two
days later, with a deep minimum in-between the two maxima. Since
UZ\,Tau is composed of two binaries, the telescope was always centered
on the spectroscopic binary UZ\,Tau\,E. However, UZ\,Tau\,W, at
a distance of 3.8$''$, also had a contribution to the measured
flux. As discussed later, we can conclude that UZ\,Tau\,W is not
brighter than UZ\,Tau\,E, and that its millimeter flux is constant in
time. Thus, we attribute the observed variability to UZ\,Tau\,E. To
rule out any instrumental artifacts (due to severe winds during the
observing run), we re-checked the pointing, which we found to be good
to within 2$''$ during all 7 UZ\,Tau observations. Therefore, the
absolute flux calibration for UZ\,Tau is precise to within 15\%, which
is better than the reported value for the entire sample. To produce a
factor of 2 \emph{decrease} in the flux---since a mispointing can only
reduce the measured flux---the pointing error would have to be 5$''$,
which is clearly not the case. Moreover, the highest fluxes we
observed are higher than values reported elsewhere, which cannot be
created by mispointing. Thus, we take the flux changes reported here
and in Fig.~\ref{fig:light_ecc} to be real.


\section{Discussion}
\label{discussion}


\subsection{Event statistics}
\label{statistics}

Our goal was to understand the proliferation of millimeter flare
events and variability within the context of a model for colliding
magnetospheres, a phenomenon that on the surface appears fairly
regular, is well described by a simple geometric model for overlapping
fields, and has a relatively clear combination of parameters ($e$ and
$d_{\rm min}$) that are predicted to lead to powerful reconnection
events visible at millimeter wavelengths. The idea is that the two
stars at apoastron possess strong independent magnetospheres, but near
periastron, the fields are more inclined to merge, or at least stretch
or compress in the presence of one another, leading to reconnection
events. This interaction radius is typically estimated to be
5\,$R_\star$ based on a dipole magnetic topology and inner disk
truncation models. However, in a binary system with a dynamically
cleared inner disk and reduced magnetic braking of the stellar
rotation from star-disk field lines, the fields may become stronger
and extend further outwards. In our subsample of high-eccentricity
candidate systems, we observed 1/6 sources to be active in the
millimeter; and 0/6 sources for the circular orbits. This brings the
number of documented millimeter-variable binaries to 3 (or 21\%) of
the combined 14-source sample, and includes V773\,Tau\,A, DQ\,Tau, and
now UZ\,Tau\,E (see Sect.~\ref{uztaue}).

We begin our analysis with the high-eccentricity sources. To proceed,
we must make use of several derived quantities from the DQ\,Tau study
to establish some detection constraints by generalizing all flares due
to interbinary, large-scale magnetospheric collision events
\citep{salter2010}. The borrowed properties include: the flare
duration ($\sim$30\,hours), peak brightness (100--500\,mJy at 3\,mm),
occurrence ($\sim$2 events per orbital period), and decay time
($\sim$6.5\,hours). All of our sources were observed approximately
twice per night over a 4-day period, with a maximum average gap in the
sampling of 16\,hours. For DQ\,Tau, a typical event is estimated to
maintain a flux $>$2 times quiescence for $\sim$75\% of a flare (or
$\sim$23\,hours for the largest outbursts), meaning that our two
high-eccentricity sources located in Taurus (140-160\,pc away) were
reasonably well sampled. However, for the four sources located in
Orion (440-470\,pc), at a distance 3 times further away where the
flux density falls off as $D^{-2}$, our detection limit is reduced by
a factor 9, meaning that a large flare may have only been detectable
for about 5-6\,hours. In these systems, flares might have been missed
by the sparse sampling. This effect might explain the non-detection
toward Parenago\,2494, which is remarkably similar, in terms of
orbital parameters, to UZ\,Tau\,E.

For the circular binaries, statistics are more difficult to discern
since the phenomenon is based on chance collisions of extended coronal
structures. Stars that are more magnetically active are likely, in
theory, to exhibit statistically more common events, but this requires
a large sample and an uninterrupted, long-duration monitoring
program. If, instead, we consider the once-a-week statistical
occurrence of giant X-ray outbursts (assuming these are magnetically
driven), then having monitored the circular sources for half a week,
we might have expected half of (all) the sources to have flared;
instead, we detected no flares toward the circular binaries. Again,
distance effects and sampling may have affected the five most distant
($>$\,400\,pc) sources.

Our study highlights the challenges of establishing a statistically
significant sample of well-parameterized spectroscopic binaries and
carrying out sufficient monitoring of the sample. In
Sect.~\ref{othersources}, we discuss the systems on a source-by-source
basis and we are able to attribute many system and stellar properties
to our non-detection statistics, including: the sensitivity of our
observations (which is a combination of a flare's flux density, the
source distance, and our 3$\sigma$ detection threshold), binary
separation, orbital eccentricity, duration of the flare, spectral peak
wavelength of the synchrotron emission, stellar magnetic field
strength and activity levels (also a function of spectral type),
magnetosphere topology, and flare decay time (a function of field
strength and separation distance). These factors may act to reduce our
detection statistics, but not the actual rate of occurrence.

In the combined 14-source sample, it is the 3 most extremely
eccentric, close-separation binaries that seem to experience very
powerful magnetic events. Therefore, while a well-defined subclass of
millimeter-flaring binaries may be taking shape, we also cannot
exclude the detected flares as a potentially rare phenomenon amongst
high eccentricity binaries.


\subsection{UZ\,Tau}
\label{uztaue}

UZ\,Tau is a hierarchical quadruple system consisting of two binary
systems UZ\,Tau\,E and UZ\,Tau\,W. The E component is a spectroscopic
binary with an orbital period of 19.1\,days \citep{mathieu1996}, and
at an angular separation of 3.8$''$ is the W component, which is
itself a 0.34$''$ binary \citep[][and references
  therein]{simon1995,prato2002}. For the primary component of
UZ\,Tau\,E, \citet{jensen2007} give a spectral type of M1 and a
photometric radius of 1.9\,$R_{\odot}$. The UZ\,Tau\,E system is a
CTTS system with ongoing accretion and clear infrared excess,
indicating that warm material can be found close to the stars
\citep[see e.g.][]{furlan2006}.

\paragraph{The millimeter picture.} At millimeter wavelengths,
\citet{simon1992} observed UZ\,Tau with the IRAM PdBI in continuum at
2.7\,mm and found that the resolved E and W components had equal
fluxes of 13$\pm$1\,mJy. Later, \citet{dutrey1996} observed the system
again with the IRAM PdBI at the same wavelength, only to report that
the E component had a flux of 25.5$\pm$1.6\,mJy, this time about 3
times brighter than the W component, which measured a more consistent
8$\pm$2\,mJy. These contrasting observations suggest that the E
component is highly variable, whereas W is more or less constant. We
note that in both papers, data obtained at different orbital phases of
the E component were combined to obtain one single flux value.

In our own survey, the UZ\,Tau source flux varied between 107 and
199\,mJy, over a period of 4 days, and between the orbital phases of
$-$0.07 and 0.09. Although the IRAM 30\,m telescope was centered on
UZ\,Tau\,E, we did not completely resolve the system. Therefore, the
W component had some contribution to these values ($\approx$ 70\% of
its flux may be included in the beam). Thus, we conclude that
UZ\,Tau\,E varied between approximately 87 and 179\,mJy at
1.25\,mm. In comparison, previous 1.3\,mm resolved fluxes from
\citet{jensen1996} and \citet{isella2009} are respectively:
137$\pm$28\,mJy and 126$\pm$12\,mJy for UZ\,Tau\,E, and 32$\pm$9\,mJy
and 30$\pm$8\,mJy for UZ\,Tau\,W. The sum of these are roughly
centered between our most extreme values. These 1.3\,mm flux values
are very similar given the earlier 2.7\,mm variability observed and
the 13-year gap in time.

If we quickly compare these multi-wavelength fluxes for UZ\,Tau\,W,
then for an average $F_{\rm 1.3mm}$\,$\approx$\,31\,mJy and $F_{\rm
  2.7mm}$\,$\approx$\,10.5\,mJy, we find a millimeter spectral slope
of $\alpha$\,$\approx$\,1.5, which is consistent with an optically
thin circumstellar (or circumbinary) disk. Performing a similar
analysis on UZ Tau E, with $F_{\rm 1.3mm}$\,$\approx$\,131.5\,mJy and
a $F_{\rm 2.7mm}$\,$\approx$\,20\,mJy, we derive an
$\alpha$\,$\approx$\,2.6, which is steeper but not unreasonable. We
remark that our minimum $F_{\rm 1.25mm}$ of 107\,mJy (or
$\approx$\,87\,mJy without the W component) gives an
$\alpha$\,$\approx$\,1.9 that is also consistent with an optically
thin disk, as well as the apparent dust evolutionary state of its
neighboring W component, which likely formed at the same time and in
the same local environment. Of course, due to the different geometry
and system parameters, they might have evolved differently. In the
end, what is most apparent, is how UZ\,Tau\,E seems to be
characterized more often than not by excess non-thermal flux. Given
the observations to date, our minimum UZ\,Tau\,E observation of
87\,mJy represents our best measurement for the true quiescent thermal
dust emission at 1.25\,mm.

The example of UZ\,Tau\,E shows that large variations in millimeter
brightness do occur, but they may have many explanations. UZ\,Tau\,E
was perhaps the most promising binary in the initial list because of
its many shared traits with fellow spectroscopic binary DQ\,Tau. This
includes of course the highest eccentricity in the sample and a close
periastron approach, but also bipolar outflows and episodic accretion
bursts from its own circumbinary disk
\citep{basri1997,jensen2007,hirth1997}.

Within the context of the DQ\,Tau scenario for colliding
magnetospheres, the apparent double-brightening present in the
UZ\,Tau\,E millimeter light curve must represent two sequential events
occurring two days apart. \citet{salter2010} also captured a secondary
brightening toward DQ\,Tau within an estimated 15\,hours of the
initial outburst. In fact, there the authors predict a minimum of two
events per periastron encounter in a scenario for colliding
magnetospheres, corresponding to first the joining, and then the
separation, of the two magnetospheres; a phenomenon that their
restricted monitoring window was unable to test fully. In this work,
covering a period of 4 days (or an observing window 6 times broader),
we find it reasonable that the first break might occur more suddenly
and closer to periastron, while the timing of the second event is
likely to occur at a much larger separation distance as the
magnetospheres slowly stretch and pinch off \citep[see Fig.~8
  in][]{salter2010}. This would then result in an asymmetry of events
around periastron, just like in our UZ\,Tau\,E light curve.

The two brightenings toward UZ\,Tau\,E possess similar recorded
maximum fluxes, unlike the sequential events captured toward DQ\,Tau
where the secondary is less bright. However, our coarse sampling means
that in all likelihood we have missed the true maxima. Our first
observed peak toward UZ\,Tau occurs when the E component stars are
separated by $\sim$22\,$R_\odot$, and the second peak occurs when they
are at a distance of $\sim$27\,$R_\odot$. The corresponding
half-separation distances of 5.8\,$R_\star$ and 7.1$R_\star$,
respectively, are consistent with an interbinary reconnection
scenario, but do not rule out other emission scenarios, for which new,
interferometric observations are necessary.

\paragraph{The optical picture.} \citet{jensen2007} obtained BVRI
photometry of UZ\,Tau\,E between 2003 and 2006, and claimed that its
light curve is periodic on timescales equivalent to the orbital
period. They interpreted these results as pulsed accretion from a
circumbinary disk, according to the model by \citet{artymowicz1996};
an accretion model that, incidentally, was first tested
observationally on data from DQ\,Tau \citep[see][]{mathieu1997}. In
this model, the circumbinary disk is periodically perturbed by the
eccentric binary components, causing the disk material to cross the
gap between the disk and the stars and fall onto the stellar
surfaces. The model predicts a smoothly varying accretion rate which
can be more or less strongly peaked at periastron depending on the
orbital parameters of the binary, and fits well the light curve
presented in \citet{jensen2007}. They note, however, that the
significant scatter in the light curves indicate that pulsed accretion
does not occur during every binary orbit. Based on arguments
concerning the amplitude and timescale of the variability,
\citet{jensen2007} discarded the possibility of rotational modulation
due to stellar spots.

\begin{figure}
\centering
\includegraphics[width=\columnwidth,angle=0]{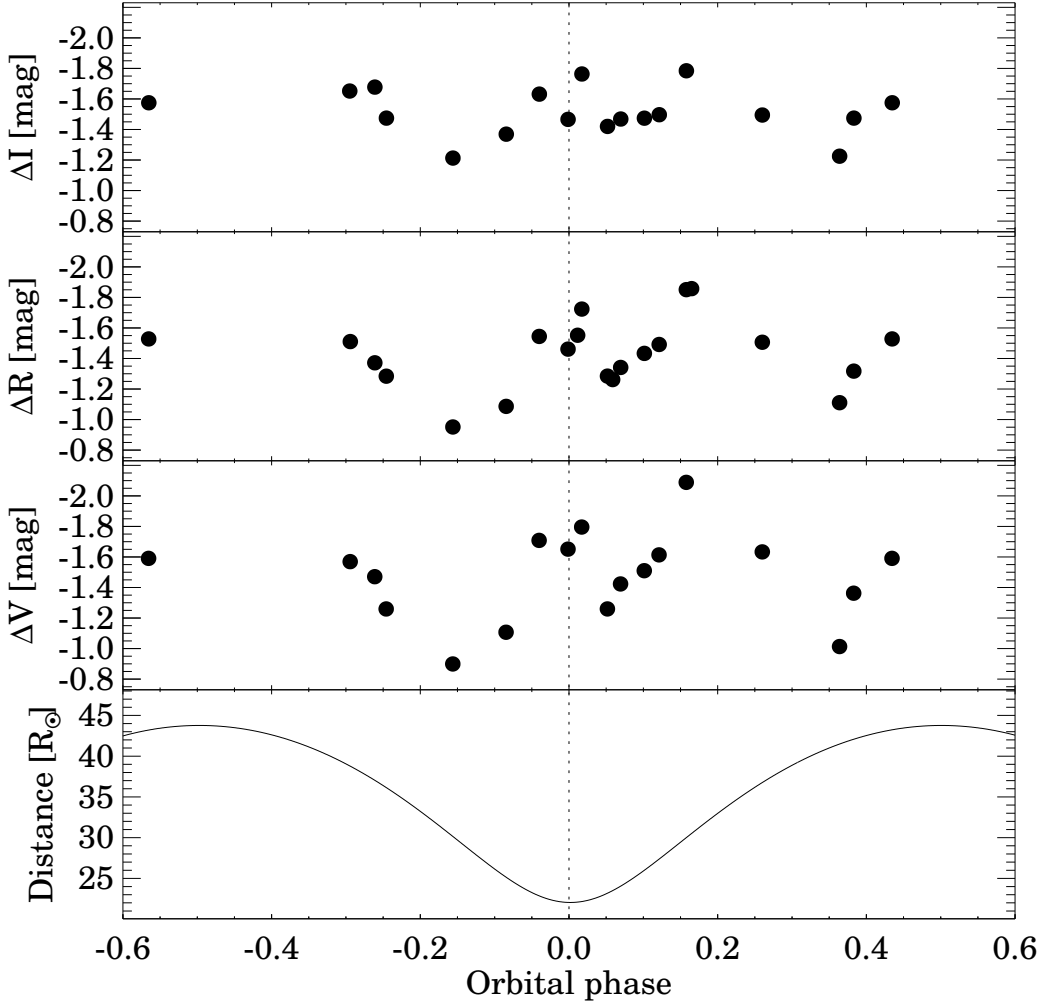}
\caption{Phase-folded optical light curves of UZ\,Tau\,E, and the
  separation of the binary components. Minima at orbital phases of
  $-$0.15, 0.05, and 0.35 are clearly visible in the light curves,
  with brightness peaks in-between. \label{fig:phase}}
\end{figure}

Our optical light curves obtained during the time of the millimeter
observations indicate two peaks: one at periastron and the other about
3 days later. The peak-to-peak amplitude is 1.19, 0.96, and 0.57 mag
in the V, R, and I band, respectively indicating an amplitude
decreasing with increasing wavelength. Additional optical observations
obtained during the previous periastron show a very similar
double-peaked light curve shape, suggesting that -- at least for two
adjacent periastrons -- the optical light curve is periodic.
Fig.~\ref{fig:phase} shows the optical light curves folded with the
orbital period of 19.131\,days. The most pronounced features in the
curves are two deep minima at orbital phases of $-$0.15 and 0.35, and
a shallower one at around 0.05. Although the light curves are not very
well sampled far from periastron, it is evident that the source was
not quiescent even when the binary components were far from each
other.

Our results indicate that apart from varying accretion rate, a
secondary contribution to the brightness changes can also be
considered. The similar shape of the millimeter and the optical light
curves suggest the possibility that optical flux changes may also be
fueled by strong magnetic activity.

\subsection{Other sources}
\label{othersources}

In the following paragraphs, we briefly discuss our results for the
rest of the binaries within the context of the available literature
data and our current analysis of the proliferation of powerful
star-star magnetic reconnection events.

\paragraph{EK\,Cep} is an eclipsing binary consisting of an A1.5V
primary and a G5Vp secondary, with radii of 1.58 and
1.31\,R$_{\odot}$, respectively \citep{torres2010}. While the primary
is already in the main sequence phase, the secondary is still a PMS
star \citep{marques2004}. When referencing published photometry and
the AKARI/IRC mid-IR all-sky survey \citep{akari}, we found no
evidence for infrared excess emission that is typically associated
with ongoing accretion or warm inner circumstellar dust. Photometric
observations by \citet{antonyuk2009} indicate that, outside of the
eclipse, no trends are noticeable in the light curves, although the
scatter of data points does exceed their photometric uncertainty
($\pm$0.04\,mag). \citet{antonyuk2009} detected variable polarization,
which they attribute to surface magnetic activity of the secondary
component. Our optical photometry shows that EK\,Cep was constant
within our measurement uncertainties ($\pm$0.02\,mag), and the source
was not detected during our millimeter observations. Our millimeter
non-detection offers an upper limit of $M_{\rm disk} \leq
0.003\,M_\odot$ for the amount of material in the cold outer disk, if
we use the basic flux-mass relation of \citet{dutrey1996}. EK\,Cep
appears to be a fairly quiescent source from optical to millimeter
wavelengths, and absent of very powerful star-star magnetic
interactions throughout an entire orbital period. This is not entirely
unexpected, since the primary component is a main sequence early
A-type star, thus it has a weak magnetic dynamo and corona, or may
completely lack these \citep{gudel2004}.

\paragraph{RX\,J0530.7$-$0434} is a WTTS binary consisting of two
identical, K2-K3 type stars with photometric radii of
3.40\,R$_{\odot}$ \citep{covino2001, marilli2007}. Optical photometry
by \citet{covino2001} revealed photometric variations with a
periodicity of 13.5 days (cf.~the orbital period of 40.57\,days). They
interpret these results by supposing that the periodicity of the light
curve is the rotation period (which is assumed to be similar for the
two binary components), indicating non-synchronous
rotation. \citet{marilli2007} give a V band amplitude of 0.22\,mag and
a rotational period of 12.9\,days. We found no evidence in the
literature for infrared excess to indicate the presence of warm
circumstellar material. Our optical light curve shows slight
variations with amplitudes of 0.12, 0.10, and 0.08 mag in V, R, and
I band, respectively, with the observed peaks approximately 5 days
before and 3 days after periastron. The source was not detected during
our millimeter observations, giving an upper limit of $M_{\rm disk}
\leq 0.024\,M_\odot$ on the disk mass.

\paragraph{Parenago\,1540} is a WTTS binary consisting of a K3V
primary and a K5V secondary with no evident infrared excess
\citep{marschall1988}. \citet{manset2002} detected photometric
variations with a V band amplitude of $<$0.5\,mag, as well as periodic
variations in the polarization, which they attribute to the orbital
motion and the fact that there is still enough dust in the environment
of the binary to produce polarization, despite the lack of infrared
excess. \citet{favata2005} reported the detection of an X-ray flare
from this source observed during the COUP survey of the Orion Nebula
Cluster. Our optical light curve shows a gradual brightening before
periastron, and a fading after periastron. The amplitude of the
variability is very similar in all bands (0.19\,mag in V, 0.18\,mag in
R, and 0.13\,mag in I). The source was not detected during our
millimeter observations ($M_{\rm disk} \leq 0.025\,M_\odot$).

\paragraph{Parenago\,2494} is a WTTS binary, the primary being an
K0IV/V star \citep{reipurth2002}. \citep{reipurth2002} found that the
star shows periodic variability in the V band with an amplitude of
0.10\,mag and a period of 5.77\,days (cf.~the orbital period of
19.48\,days). They interpret these variations as the result of large
stellar spots on the primary component (which is assumed to dominate
the optical light curve). We found no evidence in the literature for
infrared excess. Our optical light curves also show variability with
amplitudes of 0.12, 0.06, and 0.06\,mag in V, R, and I bands,
respectively. Our coverage is not enough to do a period analysis, but
the shape of the light curves is not inconsistent with a period of
about 6 days. The source was not detected during our millimeter
observations. Other than the absence of a significant circumstellar
dust reservoir ($M_{\rm disk} \leq 0.028\,M_\odot$), Parenago\,2494
possesses extremely similar orbital parameters to UZ\,Tau\,E. Located
at a distance 3 times further away, a similar outburst as the one
observed towards UZ\,Tau\,E would appear 9 times weaker, and just a
few times the noise level.

\paragraph{GG\,Ori} is a PMS eclipsing binary consisting of two nearly
identical B9.5-type stars with radii of 1.852 and 1.830\,R$_{\odot}$
\citep{torres2000}. B and V band light curves by \citet{torres2000}
display a scatter that is significantly larger than the photometric
errors, thus intrinsic variability of one or both components is
suspected. We found no evidence in the literature for infrared excess
to indicate warm material close to the stars. Our optical observations
show that GG\,Ori was constant within the measurement uncertainties
outside of the primary eclipse recorded on the night of 20/21 Nov
2009, shortly before periastron. The source was not detected at all
during our millimeter observations ($M_{\rm disk} \leq
0.023\,M_\odot$). We do remark how an eclipse that lasts longer than
the flare event itself is likely to completely obscure it, assuming
the trapped electron population lies within the eclipsing plane. In
the case of GG\,Ori, at a distance 3 times further than our outbursts
in Taurus, the detection window is also shortened as it is for
Parenago\,2494. Moreover, B-type stars are reported to be less
magnetically active than later spectral types (their magnetic fields
are below a few hundred G, and their X-ray emission is not due to
magnetospheric reconnections but shocks in the stellar wind,
\citealt{damiani1994,hubrig2009}).

\paragraph{RX\,J0350.5$-$1355} is a WTTS binary consisting of a K0-K1
primary and a K1-K2 secondary \citep{covino2001}. \citet{covino2001}
and \citet{marilli2007} found modulations in the optical light curve
with a periodicity close to the orbital period. They attribute the
0.2\,mag amplitude photometric variability to rotational modulation
due to stellar spots, suggesting that the rotational and the orbital
period is synchronized. They also note that the spectral lines of the
primary indicate faster rotation and strong magnetic activity. We
found no evidence in the literature for infrared excess. Our optical
observations show variations with amplitudes of 0.08, 0.09, and
0.12\,mag in V, R, and I band, respectively. Our coverage is not
enough to do a period analysis, but the shape of the light curves are
not inconsistent with a period of about 9 days. The source was not
detected during our millimeter observations, which cover an entire
orbital period ($M_{\rm disk} \leq 0.030\,M_\odot$). If the primary is
magnetically active, then the reconnection events may not be powerful
enough or frequent enough to be detected in the millimeter.

\paragraph{V826\,Tau} is a WTTS binary consisting of two very similar
K7V-type stars with radii of 1.44\,R$_{\odot}$
\citep{reipurth1990}. \citet{reipurth1990} found sinusoidal light
variations with an amplitude of 0.06\,mag and a period that is
slightly smaller than the orbital period. They attribute the
variability to rotational modulation due to stellar spots. We found no
evidence in the literature for infrared excess; \citet{furlan2006}
classifies it as a Class III (diskless) source. Using XMM-Newton
observations, \citet{giardino2006} detected significant X-ray
variability with a factor of 2 amplitude on a five-day timescale,
possibly related to coronal magnetic activity. Our light curves show
slight variations with amplitudes of 0.05, 0.07, and 0.04\,mag in V,
R, and I band, respectively. The source was not detected during our
millimeter observations ($M_{\rm disk} \leq 0.003\,M_\odot$), which
cover 1.5 orbital periods.

\paragraph{RX\,J0529.4$+$0041} is a WTTS triple system, consisting of
RX\,J0529.4$+$0041\,A, an eclipsing spectroscopic binary, and
RX\,J0529.4$+$0041\,B, a single star at a projected distance of
1.3$''$ \citep{covino2004}. The eclipsing binary consists of a
K1V-type primary and a K7-M0 secondary, with radii of 1.44 and
1.35\,R$_{\odot}$, respectively. \citet{covino2000} reported B and
V band brightness variations unrelated to the eclipses, possibly
connected to rotational modulation due to stellar spots and other
phenomena driven by magnetic activity. \citet{covino2004} found that
the out-of-eclipse JHK light curves of the source can be best fitted
if stellar spots are included in their model. \citet{marilli2007} also
observed out-of-eclipse variations with a V band amplitude of
0.1\,mag. Our VRI light curves show that this source was constant
within 0.03\,mag in all three bands. A possible explanation is that
the stars experienced a less active period and had no spots on their
surface during our optical observing campaign. We found no evidence in
the literature for infrared excess, and the source was undetected
during our millimeter observations ($M_{\rm disk} \leq
0.011\,M_\odot$).

\paragraph{Parenago\,1802} is a WTTS eclipsing binary consisting of
two very similar M2-type stars with radii of 1.82 and
1.69\,R$_{\odot}$ \citep{stassun2008}. \citet{cargile2008} observed
0.05\,mag peak-to-peak variations in the out-of-eclipse I band light
curve, indicating intrinsic variability. They claim that these
variations have both a periodic and a stochastic component, suggesting
that spots and chromospheric activity may both be present in the
system. Our light curves also show variability with amplitudes 0.06
and 0.05\,mag in R and I band, respectively. \citet{stassun2008}
found weak evidence for infrared excess at $>$5$\,\mu$m, indicating
the presence of a circumbinary disk and/or a faint third component in
the system. However, the source was not detected during our millimeter
observations ($M_{\rm disk} \leq 0.022\,M_\odot$).

\paragraph{RX\,J0541.4$-$0324} is a WTTS binary consisting of a G8
primary and a K3 secondary \citep{covino2001}, with photometric radii
of 2.8 and 1.8\,R$_{\odot}$, respectively. \citet{marilli2007}
observed rotational modulation in the V band light curve with an
amplitude of 0.1\,mag and period of 5 days, equal to the orbital
pediod, indicating a synchronous rotation. Our VRI light curves show
slight variations with an amplitude of 0.06\,mag in all three bands,
not inconsistent with a 5-day period. We found no evidence in the
literature for infrared excess. The source was not detected during our
millimeter observations ($M_{\rm disk} \leq 0.024\,M_\odot$).

\paragraph{NGC\,2264\,Walk\,134} is a PMS binary consisting of two
similar G-type stars with radii of 3.2\,R$_{\odot}$
\citep{padgett1994}. \citet{koch1994} reported V and R band
variability with an amplitude of 0.35\,mag, but found no periodicity
in the light curves. They also found that the V brightness of the
object is not correlated with the V$-$R color. Using Chandra
observations, \citet{ramirez2004} found that the source is variable in
the X-ray, but found no periodicity in the data, and the variability
was not flare-like either. Although its weak emission lines make it a
WTTS, it has a near-IR excess that is more typical of CTTS
\citep{padgett1994}. Archival Spitzer IRS spectra also indicates
significant infrared excess up to 30$\,\mu$m. Our optical light curves
show peak-to-peak variations of 0.22, 0.19, and 0.18\,mag in V, R,
and I band, respectively. The source appears redder when fainter. This
binary was not detected during our millimeter observations ($M_{\rm
 disk} \leq 0.088\,M_\odot$).


\section{Summary and Conclusions}
\label{summary}

Using the IRAM 30m telescope, we have conducted a monitoring program
covering 4 consecutive nights to study the millimeter variability
toward 12 PMS spectroscopic binaries mostly in the Taurus and Orion
star-forming regions. Here we report that one source, the CTTS
UZ\,Tau\,E, experiences significant millimeter flux variations
($F_{\rm 1.25\,mm}$ ranges from 87 to 179\,mJy) on daily timescales, a
clear indication of non-thermal emission processes near
periastron. The rest of the sample, consisting mainly of WTTS up to
three times more distant, remain undetected in the continuum for the
duration of the campaign, defining upper flux limits of 5--10\,mJy at
1.25\,mm (240\,GHz).

The motivation for our survey follows the recent discoveries of
recurring, bright (up to 27 times quiescent values or peaking at about
0.5\,Jy) millimeter outbursts toward the T~Tauri binaries V773\,Tau\,A
\citep{massi2008} and DQ\,Tau \citep{salter2010}. Attributed to
synchrotron activity from interbinary interactions of large magnetic
structures, the phenomenon toward V773\,Tau\,A is described as chance
collisions between extended coronal features, whereas it has been
proposed that the geometry of the DQ\,Tau system alone (specifically a
large $e$ and small $d_{\rm min}$) results in global interactions
between the two closed stellar magnetospheres near periastron
\citep{mathieu1997,basri1997}. Therefore, our target list consisted of
6 close-separation binaries with circular orbits ($e$\,$\approx$\,0)
that may experience activity at any time (but apparently did not do so
during our observing run), as well as 6 geometrically favorable
high-eccentricity systems with activity most likely to occur near
periastron (where we detect two possible events toward the source
UZ\,Tau\,E).

In our sample, no system geometry is quite as extreme as DQ\,Tau,
although our detected source UZ\,Tau\,E comes closest. Therefore, a
positive detection of (double-peaked) variability near periastron
toward UZ\,Tau\,E lends strong support to a similar global interbinary
interaction; but does not confirm it. Instead, this detection brings
our total millimeter-variable source statistics to 3
(i.e.~V773\,Tau\,A, DQ\,Tau, and UZ\,Tau\,E) out of 14 observed
sources, and means that we may need to consider that millimeter flares
are not so uncommon. In addition, as we examine the other systems in
much greater detail, it becomes clear why we might not have expected
to see, or might have missed, evidence of flares in these systems. The
most important factor seems to be the flux-distance inverse relation,
which affects our detection limits and sampling coverage. The study
itself was also limited by the number of close-separation binaries
that have been both identified and well characterized.

UZ\,Tau\,E should certainly be considered for follow-up observations
to help characterize the light curve profile, also on orbital
timescales, and to assess potential contributions from strongly
varying free-free emission processes. We must also strongly caution
against the reliability of any disk model for UZ\,Tau\,E that is based
on continuum flux measurements until the true quiescent flux level can
be established. In the future, ALMA will allow better monitoring of
these systems, leading to a more complete analysis of the
proliferation of strong millimeter activity in low-mass PMS binary
systems, as well as how much energy can be released during a
millimeter outburst and the magnetic field regeneration timescales
possible in systems known to experience recurring outbursts.


\begin{acknowledgements}
This article publishes observations made with the IAC-80 telescope,
operated by the Instituto de Astrof\'\i{}sica de Canarias at the Teide
Observatory (Spain), and with the Schmidt telescope at the
Piszk\'estet\H{o} Mountain Station of the Konkoly Observatory
(Hungary); we are grateful for the granted telescope time. We thank
C.~Zurita Espinosa for obtaining the IAC-80 data as part of a routine
observing program. We are also grateful to B.~Oca\~na and J.~Santiago
for their help during the IRAM 30\,m observations. The research of
\'A.~K., D.~M.~S., and M.~R.~H.~is supported by the Nederlands
Organization for Scientific Research (NWO). This work has benefited
from research funding from the European Community's Seventh Framework
Programme under RadioNet.
\end{acknowledgements}


\bibliographystyle{aa}
\bibliography{paper_new}{}


\Online

\begin{appendix} 
\section{Millimeter and optical data}

\begin{longtable}{ccc}
\caption{Millimeter photometry\label{tab:mm}}\\
\hline\hline
JD$-$2,450,000 & F$_{1.25\,\rm{mm}}$ [mJy] & $\sigma_{1.25\,\rm{mm}}$ [mJy] \\
\hline
\endfirsthead
\caption{continued.}\\
\hline\hline
JD$-$2,450,000 & F$_{1.25\,\rm{mm}}$ [mJy] & $\sigma_{1.25\,\rm{mm}}$ [mJy] \\
\hline
\endhead
\endfoot
\multicolumn{3}{c}{EK\,Cep}\\
\hline
5152.91 &  0.04 & 2.48 \\
5153.96 & -6.31 & 2.30 \\
5154.08 &  1.14 & 3.33 \\
5155.13 & -0.85 & 2.20 \\
5155.89 & -0.22 & 1.64 \\
5156.05 & -3.61 & 2.66 \\
\hline
\multicolumn{3}{c}{RX\,J0350.5$-$1355}\\
\hline
5153.02 & -1.13 & 2.78 \\
5154.00 &  0.47 & 2.21 \\
5154.12 & -0.86 & 3.47 \\
5155.14 & -1.21 & 1.94 \\
5155.93 & -1.82 & 2.03 \\
5156.09 &  4.86 & 2.54 \\
\hline
\multicolumn{3}{c}{V826\,Tau}\\
\hline
5152.93 & -1.16 & 2.51 \\
5153.98 &  0.46 & 1.94 \\
5154.15 & -2.02 & 2.12 \\
5155.23 & -3.84 & 1.76 \\
5155.91 &  1.91 & 1.73 \\
5156.06 &  1.20 & 1.79 \\
5156.22 & -2.82 & 3.56 \\
\hline
\multicolumn{3}{c}{UZ\,Tau}\\
\hline
5152.93 & 174.3 & 2.7 \\
5153.99 & 181.9 & 2.0 \\
5154.15 & 154.5 & 2.0 \\
5155.24 & 107.4 & 2.0 \\
5155.91 & 172.7 & 1.6 \\
5156.07 & 198.7 & 2.0 \\
5156.23 & 125.5 & 3.3 \\
\hline
\multicolumn{3}{c}{RX\,J0529.4$+$0041}\\
\hline
5153.04 & -0.66 & 2.38 \\
5154.02 & -2.61 & 2.00 \\
5154.18 & -0.58 & 2.55 \\
5155.16 & -1.00 & 1.40 \\
5155.93 & -2.56 & 2.03 \\
5156.11 & -1.82 & 1.92 \\
\hline
\multicolumn{3}{c}{RX\,J0530.7$-$0434}\\
\hline
5153.04 &  2.74 & 2.51 \\
5154.02 &  2.78 & 2.04 \\
5154.19 &  2.23 & 2.95 \\
5155.17 &  0.81 & 1.49 \\
5155.94 &  0.79 & 2.25 \\
5156.12 & -2.16 & 1.88 \\
5156.19 &  3.67 & 2.92 \\
\hline
\multicolumn{3}{c}{Parenago\,1540}\\
\hline
5153.04 & -2.35 & 2.50 \\
5154.03 & -0.64 & 2.17 \\
5154.19 &  9.87 & 2.94 \\
5155.18 &  0.22 & 1.47 \\
5155.95 &  1.45 & 2.20 \\
5156.12 &  5.12 & 2.09 \\
5156.20 &  2.65 & 2.97 \\
\hline
\multicolumn{3}{c}{Parenago\,1802}\\
\hline
5153.05 & 0.07 & 2.33 \\
5154.05 & 4.97 & 2.05 \\
5154.20 & 4.11 & 3.08 \\
5155.19 & 4.58 & 1.60 \\
5155.98 & 1.06 & 1.93 \\
5156.13 & 10.7 & 2.2  \\
5156.20 & 11.1 & 3.0  \\
\hline
\multicolumn{3}{c}{Parenago\,2494}\\
\hline
5153.05 &  3.86 & 2.30 \\
5154.10 & -0.06 & 2.13 \\
5154.21 &  2.27 & 3.51 \\
5155.20 &  0.71 & 1.62 \\
5155.99 &  1.12 & 1.91 \\
5156.14 &  1.42 & 2.23 \\
5156.21 & -5.40 & 3.26 \\
\hline
\multicolumn{3}{c}{RX\,J0541.4$-$0324}\\
\hline
5153.05 &  0.14 & 2.25 \\
5154.11 & -1.88 & 2.06 \\
5154.22 &  0.84 & 3.29 \\
5155.21 &  0.25 & 1.56 \\
5156.00 &  1.19 & 1.98 \\
5156.15 &  5.00 & 2.15 \\
\hline
\multicolumn{3}{c}{GG\,Ori}\\
\hline
5153.06 &  2.11 & 2.22 \\
5154.11 & -2.09 & 2.06 \\
5154.22 & -5.52 & 3.16 \\
5155.21 & -0.81 & 1.54 \\
5156.00 & -3.06 & 1.89 \\
5156.15 & -1.94 & 2.06 \\
\hline
\multicolumn{3}{c}{NGC\,2264\,Walk\,134}\\
\hline
5153.06 & -2.41 & 2.19 \\
5154.14 &  2.51 & 2.07 \\
5154.23 & -1.56 & 2.25 \\
5155.25 &  3.18 & 1.36 \\
5156.02 & -0.99 & 1.88 \\
5156.16 & -2.00 & 2.07 \\
\hline
\end{longtable}

\begin{table}
\caption{Comparison stars used in the optical photometry\label{tab:comp}}
\begin{tabular}{ll}
\hline
\hline
Target  &  Comparison \\
\hline
EK\,Cep              & 2MASS J21402804$+$6940328 \\
RX\,J0350.5$-$1355   & 2MASS J03501856$-$1354489 \\
V826\,Tau            & 2MASS J04320358$+$1806038 \\
UZ\,Tau\,E           & 2MASS J04323023$+$2552413 \\
RX\,J0529.4$+$0041   & 2MASS J05291738$+$0042581 \\
RX\,J0530.7$-$0434   & 2MASS J05303150$-$0434536 \\
Parenago\,1540       & 2MASS J05343988$-$0526420 \\
Parenago\,1802       & 2MASS J05351235$-$0536403 \\
Parenago\,2494       & 2MASS J05370922$-$0606445 \\
RX\,J0541.4$-$0324   & 2MASS J05412862$-$0326581 \\
GG\,Ori              & 2MASS J05431553$-$0036546 \\
NGC\,2264\,Walk\,134 & 2MASS J06405783$+$0956299 \\
\hline
\end{tabular}
\end{table}

\begin{longtable}{ccccc}
\caption{Optical photometry. The magnitudes are differential
  magnitudes with respect to the comparison stars in
  Tab.~\ref{tab:comp}. S -- Konkoly Schmidt Telescope (Hungary) ; I --
  Teide IAC-80 telescope (Spain).\label{tab:opt}}\\ \hline\hline
JD$-$2,450,000 & $\Delta$V [mag] & $\Delta$R [mag] & $\Delta$I [mag] &
Tel.\\ \hline \endfirsthead
\caption{continued.}\\
\hline\hline
JD$-$2,450,000 & $\Delta$V [mag] & $\Delta$R [mag] & $\Delta$I [mag] & Tel.\\
\hline
\endhead
\endfoot
\multicolumn{5}{c}{EK\,Cep}\\
\hline
5149.28 &             & $-$4.22(4)  & $-$3.91(3)  & S \\
5151.27 & $-$4.61(6)  & $-$4.25(2)  & $-$3.89(5)  & S \\
5154.24 & $-$4.56(11) & $-$4.23(13) & $-$3.91(13) & S \\
5155.35 & $-$3.76(5)  & $-$3.44(1)  & $-$3.18(7)  & S \\
5155.41 &             & $-$4.26(4)  &             & I \\
5156.34 &             & $-$4.22(4)  & $-$3.89(2)  & S \\
5157.38 & $-$4.62(6)  & $-$4.22     & $-$3.89(3)  & S \\
5159.24 & $-$4.61(4)  & $-$4.21(2)  & $-$3.90(7)  & S \\
5161.26 & $-$4.61(8)  & $-$4.28     & $-$3.96(5)  & S \\
\hline
\multicolumn{5}{c}{RX\,J0350.5$-$1355}\\
\hline
5149.42 & $-$3.82(1)  & $-$3.84(10) & $-$4.03(7)  & S \\
5151.47 & $-$3.75(11) & $-$3.84(1)  & $-$3.88(28) & S \\
5154.37 & $-$3.77(5)  & $-$3.84(1)  & $-$3.98(4)  & S \\
5155.44 & $-$3.81(1)  & $-$3.89(2)  & $-$4.00(1)  & S \\
5155.50 &             & $-$3.92(2)  &             & I \\
5156.40 & $-$3.80(1)  & $-$3.89(1)  & $-$3.99(2)  & S \\
5157.45 & $-$3.82(1)  & $-$3.93(1)  & $-$4.01(1)  & S \\
5157.52 &             & $-$3.92(1)  &             & I \\
5159.37 & $-$3.80(5)  & $-$3.86(1)  & $-$3.98(3)  & S \\
5161.44 & $-$3.76(6)  & $-$3.88(1)  & $-$3.91(6)  & S \\
\hline
\multicolumn{5}{c}{V826\,Tau}\\
\hline
5149.40 & $-$3.18(6) & $-$2.65(3) & $-$2.16(3) & S \\
5151.38 & $-$3.21(4) & $-$2.70(1) & $-$2.15(1) & S \\
5154.35 & $-$3.24(3) & $-$2.70(3) & $-$2.17(3) & S \\
5154.59 &            & $-$2.69(1) &            & I \\
5155.40 & $-$3.20(1) & $-$2.69(1) & $-$2.16(1) & S \\
5155.50 &            & $-$2.67(2) &            & I \\
5156.36 & $-$3.17(1) & $-$2.66(1) & $-$2.14(1) & S \\
5157.42 & $-$3.20(1) & $-$2.68(1) & $-$2.16(1) & S \\
5159.35 & $-$3.27(4) & $-$2.70(2) & $-$2.17(1) & S \\
5161.36 & $-$3.25(8) & $-$2.71(2) & $-$2.16(3) & S \\
\hline
\multicolumn{5}{c}{UZ\,Tau\,E}\\
\hline
5129.62 & $-$1.57(1)  & $-$1.51(1) & $-$1.65(1) & I \\
5130.55 & $-$1.26(1)  & $-$1.28(1) & $-$1.47(3) & I \\
5133.63 & $-$1.11(1)  & $-$1.09(1) & $-$1.37(3) & I \\
5134.48 & $-$1.71(1)  & $-$1.54(3) & $-$1.63(1) & I \\
5135.58 & $-$1.80(1)  & $-$1.72(7) & $-$1.76(1) & I \\
5136.58 & $-$1.42(4)  & $-$1.34(2) & $-$1.47(4) & I \\
5137.57 & $-$1.61(1)  & $-$1.49(1) & $-$1.50(3) & I \\
5142.58 & $-$1.36(1)  & $-$1.32(1) & $-$1.47(4) & I \\
5143.57 & $-$1.59(1)  & $-$1.53(1) & $-$1.57(1) & I \\
5149.38 & $-$1.47(36) & $-$1.37(9) & $-$1.68(2) & S \\
5151.39 & $-$0.90(5)  & $-$0.95(2) & $-$1.21(3) & S \\
5154.36 & $-$1.65(2)  & $-$1.46(2) & $-$1.47(1) & S \\
5154.60 &             & $-$1.55(1) &            & I \\
5155.37 & $-$1.26(2)  & $-$1.28(1) & $-$1.42(1) & S \\
5155.50 &             & $-$1.26(1) &            & I \\
5156.32 & $-$1.51(1)  & $-$1.43(1) & $-$1.47(1) & S \\
5157.40 & $-$2.09(3)  & $-$1.85(1) & $-$1.78(1) & S \\
5157.54 &             & $-$1.86(1) &            & I \\
5159.36 & $-$1.63(7)  & $-$1.51(2) & $-$1.49(1) & S \\
5161.34 & $-$1.01(7)  & $-$1.11(3) & $-$1.23(2) & S \\
\hline
\multicolumn{5}{c}{UZ\,Tau\,W}\\
\hline
5129.62 & $-$1.21(1) & $-$1.22(1) & $-$1.60(1) & I \\
5130.55 & $-$1.17(1) & $-$1.21(1) & $-$1.56(1) & I \\
5133.63 & $-$1.27(1) & $-$1.27(1) & $-$1.66(3) & I \\
5134.48 & $-$1.27(6) & $-$1.24(2) & $-$1.59(3) & I \\
5135.58 & $-$1.25(2) & $-$1.34(8) & $-$1.65(2) & I \\
5136.58 & $-$1.30(5) & $-$1.32(3) & $-$1.70(2) & I \\
5137.57 & $-$1.27(1) & $-$1.29(1) & $-$1.62(1) & I \\
5142.58 & $-$1.19(1) & $-$1.22(1) & $-$1.60(3) & I \\
5143.57 & $-$1.22(1) & $-$1.26(1) & $-$1.59(1) & I \\
5154.60 &            & $-$1.16(1) &            & I \\
5155.50 &            & $-$1.25(1) &            & I \\
5157.54 &            & $-$1.17(1) &            & I \\
\hline
\multicolumn{5}{c}{RX\,J0529.4$+$0041}\\
\hline
5149.44 & 0.64(2) & 0.37(2) & 0.12(2) & S \\
5151.41 & 0.65(3) & 0.36(2) & 0.10(1) & S \\
5154.38 & 0.62(2) & 0.34(2) & 0.09(1) & S \\
5154.61 &         & 0.33(1) &         & I \\
5155.42 & 0.64(1) & 0.37(1) & 0.10(1) & S \\
5155.51 &         & 0.37(1) &         & I \\
5156.38 & 0.62(1) & 0.35(1) & 0.09(1) & S \\
5157.43 & 0.64(1) & 0.36(1) & 0.11(1) & S \\
5157.54 &         & 0.36(1) &         & I \\
5159.38 & 0.61(1) & 0.34(1) & 0.08(1) & S \\
5161.45 & 0.63(1) & 0.36(5) & 0.10(1) & S \\
\hline
\multicolumn{5}{c}{RX\,J0530.7$-$0434}\\
\hline
5149.45 & $-$1.05(2) & $-$0.53(3) & 0.10(2) & S \\
5151.42 & $-$1.12(1) & $-$0.56(1) & 0.07(1) & S \\
5154.41 & $-$1.02(1) & $-$0.46(1) & 0.12(1) & S \\
5154.62 &            & $-$0.46(1) &         & I \\
5155.48 & $-$1.00(1) & $-$0.46(1) & 0.15(1) & S \\
5155.51 &            & $-$0.46(1) &         & I \\
5156.44 & $-$1.01(1) & $-$0.46(1) & 0.14(1) & S \\
5157.48 & $-$1.04(1) & $-$0.50(1) & 0.11(1) & S \\
5157.54 &            & $-$0.50(1) &         & I \\
5159.39 & $-$1.06(1) & $-$0.51(1) & 0.10(1) & S \\
5161.46 & $-$1.03(1) & $-$0.50(1) & 0.12(2) & S \\
\hline
\multicolumn{5}{c}{Parenago\,1540}\\
\hline
5149.51 & -0.92(1) & $-$1.08(1) & $-$1.19(1) & S \\
5153.49 & -1.04(7) & $-$1.13(1) &            & S \\
5154.44 & -1.02(1) & $-$1.16(1) & $-$1.26(2) & S \\
5155.52 &          & $-$1.15(1) &            & I \\
5155.60 & -1.02(1) & $-$1.17(1) & $-$1.26(3) & S \\
5156.55 & -1.06(1) & $-$1.20(1) & $-$1.29(1) & S \\
5157.56 & -1.11(1) & $-$1.25(1) & $-$1.32(2) & S \\
5161.38 & -1.08(1) & $-$1.19(1) & $-$1.30(5) & S \\
\hline
\multicolumn{5}{c}{Parenago\,1802}\\
\hline
5149.52 &          & 0.60(1)    & $-$0.26(2) & S \\
5154.43 &          & 0.56(1)    & $-$0.30(1) & S \\
5154.65 &          & 0.61(2)    &            & I \\
5155.52 &          & 0.60(1)    &            & I \\
5155.58 &          & 0.59(1)    & $-$0.27(1) & S \\
5156.52 &          & 0.61(1)    & $-$0.26(1) & S \\
5157.55 &          & 0.57(1)    & $-$0.28(1) & S \\
5157.55 &          & 0.58(1)    &            & I \\
5161.37 &          & 0.62(1)    & $-$0.25(1) & S \\
\hline
\multicolumn{5}{c}{Parenago\,2494}\\
\hline
5149.50 & $-$2.78(1) & $-$2.52(1) & $-$2.36(1) & S \\
5153.47 & $-$2.70(2) & $-$2.49(1) & $-$2.34(1) & S \\
5154.42 & $-$2.73(3) & $-$2.50(1) & $-$2.35(3) & S \\
5154.66 &            & $-$2.54(1) &            & I \\
5155.50 & $-$2.78(1) & $-$2.53(1) & $-$2.37(1) & S \\
5155.53 &            & $-$2.52(1) &            & I \\
5156.46 & $-$2.82(1) & $-$2.56(1) & $-$2.40(1) & S \\
5157.50 & $-$2.77(2) & $-$2.53(2) & $-$2.39(1) & S \\
5157.56 &            & $-$2.52(2) &            & I \\
5161.39 & $-$2.79(2) & $-$2.52(2) & $-$2.39(4) & S \\
\hline
\multicolumn{5}{c}{RX\,J0541.4$-$0324}\\
\hline
5149.49 & $-$0.60(1) & $-$0.76(2) & $-$0.90(2) & S \\
5151.44 & $-$0.55(1) & $-$0.70(2) & $-$0.84(1) & S \\
5154.40 & $-$0.59(1) & $-$0.73(4) & $-$0.90(2) & S \\
5154.67 &            & $-$0.72(1) &            & I \\
5155.46 & $-$0.58(1) & $-$0.73(1) & $-$0.88(1) & S \\
5155.53 &            & $-$0.72(1) &            & I \\
5156.42 & $-$0.54(1) & $-$0.70(1) & $-$0.85(1) & S \\
5157.47 & $-$0.57(1) & $-$0.72(1) & $-$0.86(1) & S \\
5157.56 &            & $-$0.71(1) &            & I \\
5159.47 & $-$0.60(1) & $-$0.76(1) & $-$0.88(1) & S \\
5161.47 & $-$0.55(1) & $-$0.70(1) & $-$0.84(1) & S \\
\hline
\multicolumn{5}{c}{GG\,Ori}\\
\hline
5151.48 &            & $-$0.10(2) & $-$0.06(3) & S \\
5153.44 & $-$0.31(1) & $-$0.13(2) & $-$0.05(1) & S \\
5154.67 &            & $-$0.10(1) &            & I \\
5155.54 &            &    0.19(1) &            & I \\
5155.54 & $-$0.03(3) &    0.15(3) &    0.22(2) & S \\
5156.49 & $-$0.31(1) & $-$0.12(1) & $-$0.05(1) & S \\
5157.52 & $-$0.30(1) & $-$0.11(1) & $-$0.05(1) & S \\
5157.55 &            & $-$0.11(1) &            & I \\
5159.48 & $-$0.29(1) & $-$0.10(1) & $-$0.01(1) & S \\
5161.42 & $-$0.29(2) & $-$0.10(2) & $-$0.04(1) & S \\
\hline
\multicolumn{5}{c}{NGC\,2264\,Walk\,134}\\
\hline
5149.53 &    0.10(1) & $-$0.07(1) & $-$0.23(1) & S \\
5151.45 & $-$0.01(1) & $-$0.15(1) & $-$0.31(1) & S \\
5155.52 &    0.15(1) & $-$0.01(1) & $-$0.19(1) & S \\
5155.54 &            & $-$0.02(1) &            & I \\
5156.48 &    0.22(1) &    0.04(1) & $-$0.13(1) & S \\
5157.51 &    0.04(1) & $-$0.12(1) & $-$0.28(1) & S \\
5157.51 &            & $-$0.12(1) &            & I \\
5161.40 &    0.02(1) & $-$0.14(1) & $-$0.29(1) & S \\
\hline
\end{longtable}

\end{appendix}

\end{document}